\documentclass[%
reprint,
superscriptaddress,
 amsmath,amssymb,
 prr,
floatfix,
]{revtex4-2}

\usepackage{graphicx}% Include figure files
\usepackage{dcolumn}% Align table columns on decimal point
\usepackage{bm}% bold math
\usepackage{hyperref}% add hypertext capabilities
\hypersetup{
    colorlinks = true,
    allcolors=black,
    citecolor=blue,
    linkcolor=blue
} % add hypertext color
\usepackage{amsmath, amsfonts, dsfont, mathrsfs, float,amssymb}
\usepackage{color}
\newcommand{\trace}[2]{\text{\normalfont Tr}_{#1} \left[#2\right]}
\newcommand{\barj}[1]{\bar{\jmath}}
\begin{document}

%\preprint{APS/123-QED}

\title{A Memory Hierarchy for Many-Body Localization: Emulating the Thermodynamic Limit}
%\title{Memory Hierarchy and Artifical Decoherence across the Many-Body Localization Transition}

\author{Alexander Nico-Katz}
\email{alex.nico-katz.15@ucl.ac.uk}
\affiliation{Department of Physics and Astronomy, University College London, London WC1E 6BT, United Kingdom}
\author{Abolfazl Bayat}
\email{abolfazl.bayat@uestc.edu.cn}
\affiliation{Institute of Fundamental and Frontier Sciences, University of Electronic Science and Technology of China, Chengdu 610051, China}
\author{Sougato Bose}
\email{s.bose@ucl.ac.uk}
\affiliation{Department of Physics and Astronomy, University College London, London WC1E 6BT, United Kingdom}

\date{\today}% It is always \today, today,
             
\begin{abstract}
    Local memory - the ability to extract information from a subsystem about its initial state - is a central feature of many-body localization. We introduce, investigate, and compare several information-theoretic quantifications of memory and discover a hierarchical relationship among them. %This hierarchy is bounded on both sides by  widely-used na\"ive memory quantifiers, namely von Neumann entropy and imbalance. 
    We also find that while the Holevo quantity is the most complete quantifier of memory, vastly outperforming the imbalance, its decohered counterpart is significantly better at capturing the critical properties of the many-body localization transition at small system sizes. This motivates our suggestion that one can emulate the thermodynamic limit by artifically decohering otherwise quantum quantities. Applying this method to the von Neumann entropy results in critical exponents consistent with analytic predictions, a feature missing from similar small finite-size system treatments. In addition, the decohering process makes experiments significantly simpler by avoiding quantum state tomography.
\end{abstract}
             
\maketitle

%\tableofcontents

\section{Introduction}
\label{sec:intro}
Many-body localization (MBL) in disordered systems has been the subject of varied theoretical and experimental studies~\cite{Anderson1958, Basko2006, Imbrie2016, Nandkishore2014, Altman2015, Alet2018, Abanin2019} and its features are widely recognised: (i) strong ergodicity-breaking and ETH violation in the MBL phase \cite{Deutsch1991, Srednicki1994, Rigol2008}, (ii) a rich many-body localization phase transition (MBLT) across the entire spectrum with highly debated properties \cite{Pal2010, grover2014certain, Luitz2015, Potter2015, Khemani2017, Khemani2017a, Thiery2018, Goremykina2019, Dumitrescu2019, Morningstar2020}, and (iii) emergent integrability and associated local memory \cite{Serbyn2013, Huse2014, Ros2015, Chandran2015liom, Imbrie2016} which provides a lucrative testbed for applications in quantum computing. Most of these characteristic signatures of MBL have been placed on firm quantitative footings through the imbalance~\cite{Schreiber2015, Luitz2016, Iemini2016, Levi2016, Choi2016, Luschen2017} and magnetization \cite{Pal2010, Luitz2015, Singh2016, Sierant2017}, entanglement entropies~\cite{Bardarson2012, Herviou2012, Serbyn2013entanglement, Kjall2014, Luitz2016, Singh2016}, spatial correlations~\cite{Bera2016, Detomasi2017}, unfolded level statistics~\cite{Oganesyan2007, Pal2010, Herviou2012, Ponte2015, Luitz2015, De2021}, and myriad others~\cite{Bera2016, Iemini2016, Gray2018, Gong2021}. A key concept in MBL is the emergence of memory: the local retention of information about initial conditions. Memory is necessarily temporal and thus typically discussed through dynamical quantities such as the imbalance, steady-states of local observables, or deduced from entanglement and correlation spreading~\cite{Ponte2015, Luitz2016, Levi2016, Bordia2016, Gullans2020, Kvorning2021}. However, these quantities suffer from a range of problems which curtail their use as a true quantifier of memory. In fact, a good quantifier of memory is expected to satisfy two criteria: (i) it should have an informational interpretation in terms of the number of bits and (ii) it should contain temporal correlations.
To satisfy these criteria, an informational framework and in particular the Holevo quantity - an upper bound on the amount of classical information which a quantum channel can bear - have recently been introduced in MBL systems~\cite{NicoKatz2020}.\begin{figure}[ht]
        \includegraphics[width=\linewidth]{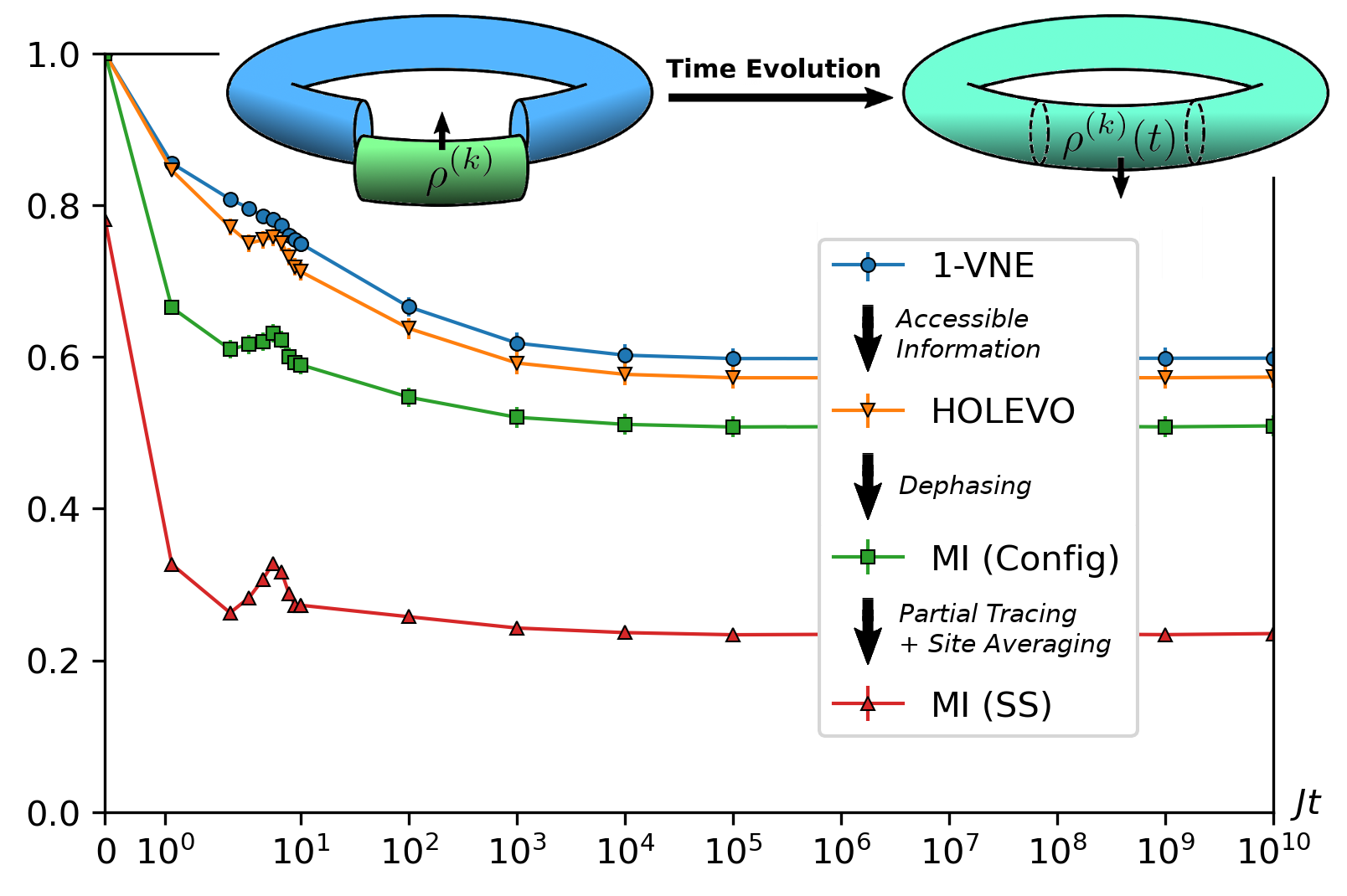}
        \caption{Hierarchy of disorder-averaged informational quantities against time (normalized between 0 and 1) for disorder strength $h=4$ in the middle of the MBLT. The schematic shows the action of a single channel realization on a single contiguous message state initialized in the state $\rho^{(k)} = |k\rangle\langle k|$. As each channel realization has a fixed disorder profile and environment state, we average over channels when constructing the quantities. The peak at $Jt\approx8$ occurs when the information carried away from the message first interacts with itself, causing transient revival. Error bars shown where visible, results for both sides of the MBLT, $h=0.1$ and $h=16$, are given in Appendix \ref{sec:app-hierarchy}. All results are for $L=16$, $l = 4$.}
        \label{fig:fig1}
\end{figure}
%  While many of these quantities are loosely used to capture the features of memory, a systematic approach for its quantification and comparison between these quantities is missing. In particular, an important open question is whether one can establish a hierarchy between different memory quantifiers in MBLt systems.
A systematic investigation among memory quantifiers, in accordance with the above criteria, is missing; as is a comparison of their relationship to extant quantities, especially in the context of MBL. In such an investigation, two important questions can be asked. First, which memory quantifier preserves the most information over time? Second, which memory quantifier best captures the properties of the ergodic-MBL transition? In particular, can one can establish a hierarchy on either of these grounds between different quantifiers and determine which best captures memory in MBL systems?
% lunt/etc. papers discuss diagonal entropy in terms of $p_i = |<E_i|\psi(0)\rangle|^2$, not our type of diagonal entropy:  
%https://journals.aps.org/prresearch/pdf/10.1103/PhysRevResearch.2.043072
%https://iopscience.iop.org/article/10.1209/0295-5075/101/37003/pdf

A central issue in MBL physics is the characterization of the transition point. Despite advancements in quantum simulators which allow for experimental attempts to access the transition~\cite{Choi2016, Roushan2017, Xu2018, Zha2020, guo2020stark, Guo2020, Gong2021} and theoretical breakthroughs, the nature of the transition is still highly debated~\cite{Pal2010, grover2014certain, Luitz2015, Potter2015, Khemani2017, Khemani2017a, Thiery2018, Goremykina2019, Dumitrescu2019, Morningstar2020}. 
Assuming the existence of a second order phase transition, analytical results by Harris~\cite{Harris1974} and developments thereof \cite{Chayes1986, Chandran2015} bound the critical exponent $\nu \geq 2$. %which describes the divergence of the length scale of the system across the MBL transition point.% with the exception of schmidt and sun, nu < 2 and contradict harris bound.
Nevertheless, the limitation of exact numerics to small system sizes ($L \sim 20$) makes extracting thermodynamical behavior challenging; almost all small-scale numerical analyses - with few exceptions~\cite{Gray2018, Sun2020} - violate the Harris bound. Large phenomenological approaches - though blind to microscopic features of the system - have had markedly more success: achieving this bound \cite{Potter2015, Vosk2015, Dumitrescu2017, Morningstar2019, Goremykina2019}. Recent work even suggests that the transition may be of a Kosterlitz-Thouless type, and thus the Harris bound is inapplicable \cite{Dumitrescu2019, Morningstar2020, Suntajs2020, Laflorencie2020}. %One may wonder, whether a systematic investigation of memory can shed light on the nature of the transition.
One may wonder whether a novel investigation of memory in small systems can reveal improved methods for analysing the MBLT, bringing the small scale analyses in line with analytic and phenomenological results.

% even for $L \leq 15$
The paper is organized as follows. We begin by reviewing the model and unitary evolution in section \ref{sec:model}. We then introduce, define, and discuss four different quantities in the informational framework, some of which bear resemblance to non-informational quantities already used in context of MBL, in section \ref{sec:quantities}. We compare these quantities with respect to two standards. Firstly, in section \ref{sec:compare}, which best captures memory in terms of the amount of information - in bits - retained over time; an analysis which reveals a strict hierarchy of memory quantifiers. Secondly, in section \ref{sec:scaling}, which best captures critical behaviour across the MBL transition in terms of minimizing the violation of the Harris bound~\cite{Harris1974,Chayes1986, Chandran2015}. This analysis reveals the insight that by artificially decohering a subsystem one can consistently reduce the amount by which the Harris bound is violated - in some cases even yielding results \textit{consistent} with it. We discuss this finding in \ref{sec:emulate} and suggest that this is due to the fact that artificially decohering a subsystem better emulates behavior at the thermodynamic limit, which we motivate with some preliminary results.

%In this Letter, we begin to address the above questions by investigating four related informational quantities, three of which are quantifications of memory in the sense that they are (i) temporal: they correlate the information inserted into a system in the past with the information extracted from that system's future (ii) informational: they yield results in units of information: namely the bit. One such quantity - an informational version of the widely used imbalance - serves as a benchmarking tool. We will compare them on the grounds of which serves best as a quantifier of `memory', and which best captures the MBLT in small systems $L \leq 15$ accessible to exact diagonalization and experiment alike. This comparison includes a thorough series of scaling analyses. We will show that these comparisons reveal a novel insight about dephasing in small systems undergoing the MBLT: that artifically dephasing systems yields better scaling results, some even in line with the Harris bound \cite{Harris1974,Chayes1986, Chandran2015}. We argue that wholly quantum quantities may actually be worse at capturing the MBLT in small or medium systems than their (classical) dephased variants.

\section{Model} 
\label{sec:model}
We consider a system of $l$ spin-1/2 particles which encode pure separable messages of the form $|k\rangle=|k_1,k_2,\cdots,k_l\rangle$ where $k_1 k_2 \cdots k_l$ is a binary representation of $k$, and where each $k_i \in \{0,1\}$ is encoded in the physical system by spin up $|\uparrow\rangle$ and spin down $|\downarrow\rangle$ states respectively. The index $k$ runs over all $2^l$ configurations and thus $\{|k\rangle\}$ forms a complete orthonormal basis for the system, which we call the ``message register". Note that, for the rest of this paper, we only consider contiguous message registers.  Each message $|k\rangle$ is inserted into the message register with probability $p_k$. In this paper, we consider equiprobable messages, namely $p_k=1/2^l$. The message register is then embedded in a random environment $|E\rangle$ of size $L-l$ (i.e. the total system size is $L$) which is initialized in a random product state of up and down spins. Thus, the initial state of the entire system is described by $|k\rangle\langle k|\otimes|E\rangle\langle E |$. The interactions between the particles are explained by the periodic Heisenberg Hamiltonian with random fields
\begin{equation}\label{eq:ham}
   H = J\sum_{j=1}^{L} S_j \cdot S_{j+1} +\sum_{j=1}^L h_j S^z_j
\end{equation}
where $J$ is the exchange coupling, $S_j=(S^x_j,S^y_j,S^z_j)$ is the vector of spin operators at site $j$,  and the $h_i$'s are random fields drawn uniformly in the interval $[-h,+h]$, with $h$ characterizing the disorder strength. This model has been used extensively in MBL literature, and is expected to localize for $h > h_c \gtrsim 3.7$ \cite{Khemani2017, Laflorencie2020}. 
The quantum state of the message register at any time $t$, defined as $\rho^{(k)}(t)$, is given by the unitary evolution of the entire system and the partial tracing out of the environmental degrees of freedom:
\begin{equation*}
    \rho^{(k)}(t) {=} \trace{E}{e^{-i H t} \left ( |k\rangle\langle k | \otimes |E\rangle  \langle E|  \right)  e^{i H t} }. 
\end{equation*}
This procedure is shown in the schematic of Fig.~\ref{fig:fig1}. The goal is to quantify how much information - in bits - can be extracted locally from the message register at time $t$ about its initial state. This temporal correlation between the past and future is, essentially, ``memory''.

\section{Informational Quantities}
\label{sec:quantities}
We will consider four quantities: (i) the von Neumann entropy, which has seen consistent use in MBL; (ii) the Holevo quantity; %, which can be  utilized for quantifying memory in MBL systems;
(iii)  the classical single-site mutual information (SSMI) which is closely related to the widely-used imbalance quantity; and (iv) the classical configurational mutual information (CMI), which has not been studied in the context of MBL. The last three are all true memory quantifiers in the sense that a memory quantifier is a quantity which (i) temporally correlates the initial and final states of a subsystem, and (ii) has a bitwise interpretation. Memory quantifiers can have various advantages over each other such as being optimal over measurement bases or experimentally tractable \cite{NicoKatz2020}. Whilst the schematic of Fig.~\ref{fig:fig1} shows the effect of a single channel realization, each of these four quantities is averaged over 500-1000 realizations of the channel - 
%different messages (denoted by $\sum_k p_k \cdot$)  and different choices of 
the random fields and environment states - in a process we denote by $\left\langle \cdot \right\rangle$. For a summary of all quantities, see Table~\ref{tab:table1}.

\begin{table*}[ht]
    \centering
    \renewcommand{\arraystretch}{1.5}
    \begin{tabular}{|c||c|c|c|} \hline
         \textbf{Quantity} & \textbf{Memory Quantifier} & \textbf{Measurements} & Definition \\ \hline\hline
         Modified von Neumann Entropy & No & Full State Tomography & Eq.~(\ref{eq:vne})\\
         Holevo Quantity & Yes & Full State Tomography & Eq.~(\ref{eq:Holevo_def})\\
         Configurational Mutual Information (CMI) & Yes & Global Measurements & Eq.~(\ref{eq:mutual_c})\\
         Single-Site Mutual Information (SSMI) & Yes & Local Measurements & Eq.~(\ref{eq:miss-def})\\ \hline
    \end{tabular}
    \caption{Summary of the four informational quantities discussed in this paper; whether they hold the status of a valid memory quantifier based on the criteria laid out in this paper and Ref.~\cite{NicoKatz2020}, and what measurements need to be made to construct the quantity.}
    \label{tab:table1}
\end{table*}

%We use the the bar average notation $\bar{\cdot}$ for denoting the overall average of a quantity: some combination of the above two averages. 
In the following, we provide quantitative definitions for, and intuitive explanations of, these quantities. We emphasize that, though we have made an explicit choice of equiprobable messages such that the initial distribution is uniform $p_k = 1/2^l$, these definitions generalize to any choice of $p_k$.
%Moreover, We normalize all quantities except the single-site mutual information by $1/l$ such that they all lie between zero and unity. This allows us to interpret the Holevo and mutual informational quantities as average bit rates, and the von Neumann entropy as their upper-bound.

We define the normalized von Neumann entropy of the message register as 
\begin{equation}\label{eq:vne}
    \overline{S(t)} = \frac{1}{l}\left\langle{\sum_k p_k S(\rho^{(k)}(t))}\right\rangle
\end{equation}
where $S(\rho) = -\text{Tr}\rho\log_2\rho$. One may be motivated to suggest that the more mixed the subsystem $\rho^{(k)}(t)$ is on average, the less information has been retained within said subsystem; and as such $1-\overline{S(t)}$ is a quantifier of memory. However this quantity contains no \textit{accessible} informational content: in fact, whilst $\overline{S(t)}$ quantifies the average \textit{instantaneous} mixedness of the states $\rho^{(k)}(t)$, it does so without regard for how well they bear information from the past or into the future. Consider - for instance - a channel which sends all initial states $\{|k\rangle\}$ to the single pure state $|0\rangle$. Clearly this channel is entirely useless for transmitting information, but $1-\overline{S(t)}$ still saturates to unity. As such $1-\overline{S(t)}$ is unsuitable as a memory quantifier. However, as we will see in the following, it upper-bounds all other true memory quantifiers, and is widely used in MBL literature so serves as a useful benchmark.

The Holevo quantity~\cite{holevo1973,Roga2010} is a measure of the accessible classical information which can be accessed via optimal measurements on an ensemble of quantum states. Intuitively it can be understood as the maximum classical capacity of a quantum channel \cite{nielsen2011}. In Ref.~\cite{NicoKatz2020} it has been proposed as a quantification of memory in MBL systems. For our setup, the normalized average Holevo quantity is defined as
\begin{equation}
\overline{C(t)} = \frac{1}{l} \left\langle S\left(\sum_k p_k\rho^{(k)}(t)\right) - \sum_k p_k S\left(\rho^{(k)}(t)\right) \right\rangle \label{eq:Holevo_def}
\end{equation}
where the two message averages temporally connect the initial distribution encoded in the $p_k$'s with the final distribution encoded in the mixed state of the message register $\rho^{(k)}(t)$ - the essential feature of memory. We can intuit the Holevo quantity as constructed in Eq.~(\ref{eq:Holevo_def}) as a correction to $1-\overline{S(t)}$ which accounts for whether or not the ensemble $\{\rho^{(k)}(t)\}$ actually bears information over time. Consider again the channel which sends all messages to $|k\rangle \to |0\rangle$, the first term in Eq.~(\ref{eq:Holevo_def}) accounts for this and causes $\overline{C(t)}$ to saturate to zero. Thus, whilst $1-\overline{S(t)}$ upper bounds all quantities, $C(t)$ upper bounds the actual amount of \textit{accessible} information contained in the system over time \footnote{This is true given our fixed probability distribution $p_k$, the Holevo quantity in practice is maximized over this distribution.}. 

Experimentally, it is most convenient to measure local quantities in a single measurement basis, e.g. computational basis. The joint probability that a site $j$ in the message $|k\rangle$ is initialized in the state $\sigma \in \{0, 1\}$ and then measured at time $t$ in the state $\sigma^\prime \in \{0, 1\}$ is given by:
\begin{equation}\label{eq:ss-probs-full}
    p_{k,j}(\sigma, \sigma^\prime, t) = \text{tr}\left[ P_j^\sigma |k\rangle \langle k| \right]\text{tr}\left[P_j^{\sigma^\prime}\rho^{(k)}(t)\right]
\end{equation}
where $P^\sigma_j=|\sigma_j\rangle \langle \sigma_j|$ is the projection operator at site $j$. These probabilities, when summed over all sites and states $\sigma$, $\sigma^\prime$ using appropriate phases and rescaled, is identical to the imbalance - a quantity which has been extensively utilized in MBL theory and experiment (see for example Refs.~\cite{Schreiber2015, Luitz2016, Iemini2016, Levi2016, Choi2016, Luschen2017}) as a na\"ive measure of memory. To briefly summarize: the imbalance $\mathcal{I}$ is a sum over single-site autocorrelation functions in a fixed basis e.g. for spinless fermions (the context in which much of the work on the imbalance takes place) $\mathcal{I} = \sum_j \langle \Psi | n_j(0) n_j(t) | \Psi \rangle$ where $n_j(t)$ is the Heisenberg form of the number operator for site $j$ at time $t$ and where $|\Psi\rangle$ is the initial state of the system. In the ergodic regime $\mathcal{I}$ fluctuates around zero, while in the MBL regime it fluctuates around a finite non-zero value. We can clearly see, in the case that $|\Psi\rangle$ is an eigenstate of each local number operator at $t=0$, that we can rewrite each term in $\mathcal{I}$ in a form similar to Eq.~(\ref{eq:ss-probs-full}). In essence, Eq.~(\ref{eq:ss-probs-full}) are information-theoretic versions of the kind of measurements that are used to construct the imbalance. We then define the average SSMI as a mutual information in terms of the $p_{k,j}(\sigma, \sigma^\prime, t)$
\begin{equation}\label{eq:miss-def}
    \overline{I_s(t)} = \left\langle \sum_{k, \sigma, \sigma^\prime}  p_k \tilde{p}_{k}(\sigma, \sigma^\prime, t) \log_2 \frac{\tilde{p}_{k}(\sigma, \sigma^\prime, t) }{\tilde{p}_{k, \Sigma}(\sigma, t) \tilde{p}_{k, \Sigma^\prime}(\sigma^\prime, t) }  \right\rangle 
\end{equation}
where $\tilde{p}_{k}(\sigma, \sigma^\prime, t) = \sum_j {p_{k,j}}(\sigma, \sigma^\prime, t)/l$ and where $p_{k,\Sigma(\Sigma^\prime)}(\sigma(\sigma^\prime), t) = \sum_{\sigma^\prime(\sigma)}\tilde{p}_k(\sigma, \sigma^\prime, t)$ are the marginal distributions of our site-averaged initial and final spin state variables respectively. Thus, the SSMI is a natural information-theoretic extension of the imbalance in the sense that it is the information that can be extracted from the subsystem $\rho^{(k)}(t)$ about its initial configuration using \textit{only} the measurements taken when computing the imbalance, namely single-site measurements in a fixed basis.

One can extend the above idea to the whole message register, taking measurements in basis of the message states $\{|k\rangle\}$. In this case, the outcome of the measurements is one of the $2^l$ spin configurations. The joint probability $p(k, k^\prime, t)$ of sending the state $|k\rangle$ and measuring the state $|k\rangle$ at time $t$ is given by
\begin{equation}
    p(k, k^\prime, t) = p_k \langle k^\prime | \rho^{(k)}(t) | k^\prime \rangle.
\end{equation}
We then define the normalized CMI for this joint probability as
\begin{equation}\label{eq:mutual_c}
    \overline{I_c(t)} = \frac{1}{l} \left \langle \sum_{k k^\prime} p(k, k^\prime, t) \log_2 \frac{p(k, k^\prime, t)}{ p_K(k, t) p_{K^\prime}(k^\prime, t)} \right \rangle
\end{equation}
where $p_{K(K^\prime)}(k(k^\prime), t) = \sum_{k^\prime (k)} p(k, k^\prime, t)$ are the marginal distributions of our initial and final state variables, respectively. Importantly, as shown in the SM, this construction is exactly the Holevo quantity of Eq.~(\ref{eq:Holevo_def}) in the limit of a fully decohered message register, in which $\rho^{(k)}(t)$ is replaced by a decohered density matrix $\rho^{(k)}_D(t)$ with all off-diagonal elements set to zero
\begin{equation}\label{eq:decohered}
    \rho^{(k)}(t) \to \rho_D^{(k)}(t) = \text{diag}(\rho_{11}^{(k)}(t), \rho_{22}^{(k)}(t), \cdots, \rho_{2^l 2^l}^{(k)}(t)).
\end{equation}
In this sense, we can understand $\overline{I_c(t)}$ as direct counterpart to the quantity $\overline{C(t)}$ after all output states have been passed through the fully decohering channel of Eq.~(\ref{eq:decohered}). We emphasize here that the decohering process occurs only on the message register \textit{after} full unitary evolution up to time $t$. Thus, while it destroys all coherence within the message register at the time of measurement, it doesn't preclude the build up of long-range coherences during the unitary evolution of the full system.

Finally, we draw attention to an important feature of the memory quantifiers in that they are each constructed using different kinds of measurements. The Holevo quantity requires full tomography of the subsystem, the CMI requires measurements in a fixed global basis on the subsystem, and the SSMI requires the same measurements as the imbalance: namely local measurements in a fixed basis. As such, these quantities give us a useful quantification of the maximum amount of information that can be extracted from a subsystem given the measurements accessible to the experimenter - a crucial point when considering the potential of MBL systems in quantum computational settings. We summarize all quantities, the measurements required to construct them, and their status as memory quantifiers in Table~\ref{tab:table1}.

%There is an important technical issue which is the question of when to take the message and site averages over the indices $k$ and $j$ respectively\footnote{Explicit computation of the marginal distributions over $\sigma$ and $\sigma^\prime$ makes it clear that the mutual information without performing either of the averages over $k$ and $j$ is zero - a consequence of the fact that a protocol which uses one message state can convey no meaningful information.}. We will take the average over $j$ before we compute the mutual information, and the average over $k$ afterwards (at the same time as we take the disorder and environment average). Detailed justifications for this decision, and results for taking both averages before computing the mutual information, are available in For now it suffices to say that the ordering we use here both mirrors the typical construction of the imbalance, and actually yields scaling results (albeit weak ones) - the other ordering does neither.

\section{Comparison of Memory Quantifiers}
\label{sec:compare}
The above quantities, except the normalized von Neumann entropy $\overline{S(t)}$, establish temporal correlations between the distribution of messages inserted into the message register at $t=0$, and their final states at time $t$. In addition, all these quantities are defined such that they are bounded between 0 and 1. Thus the SSMI, CMI, and Holevo quantity are directly comparable as memory quantifiers, and each capture local memory differently depending on the measurements available. The von Neumann entropy upper bounds all quantities and serves as a useful benchmark. We compare these quantities with respect to two standards. Firstly, which better captures local memory in terms of the number of bits of information preserved in the subsystem over time. Secondly, which better captures the critical properties of system Eq.~(\ref{eq:ham}) as we scan $h$ across the MBLT in terms of how strongly each quantity violates the Harris bound. The latter case is discussed in the scaling analysis later in this paper, but the former question: which quantity is a better memory quantifier, is quickly addressed here.

In Fig.~\ref{fig:fig1}, we plot these four quantities as a function of time for a system of $L=16$, $l=4$ and $h/J=4$. Exact diagonalization and subsequent time evolution were carried out using modifications of the quimb package \cite{Gray2018quimb}. One can clearly see that these quantities form a strict hierarchy at all times. Indeed, we conjecture that the four quantities obey the following inequality for all choices of $L$, $l$ and $h$
\begin{equation}\label{eq:hierarchy}
 1-\overline{S(t)} \ge \overline{C(t)} \ge \overline{I_c(t)} \ge \overline{I_s(t)}
\end{equation}
In Appendix \ref{sec:app-hierarchy}, we strictly prove that $1 - \overline{S(t)} \ge \overline{C(t)} \ge \overline{I_c(t)}$. The last part of the inequality of Eq.~(\ref{eq:hierarchy}), namely the CMI upper bounding the SSMI, has been extensively verified through numerical simulation and is intuitively sensible: dephasing and tracing out parts of a system should not increase the information we can glean from it. 
%\color{blue}{ EXCLUDE ? However, we lack this part of the proof, and must state the right hand side of the above inequality as conjecture.}\color{black}  
Indeed all the inequalities above have intuitive origins in quantum information: as discussed in the section \ref{sec:compare}, the Holevo quantity is a correction to $1-\overline{S(t)}$ that specifically addresses whether or not information in the subsystem is accessible given optimal measurements. The CMI is simply the Holevo quantity after all final message states have been passed through the fully dephasing channel in Eq.~(\ref{eq:decohered}), and by the monotonicity of the Holevo quantity under such a channel the inequality $\overline{C(t)} \ge \overline{I_c(t)}$ is evident. This is illustrated in the legend of Fig.~\ref{fig:fig1}, showing the hierarchy and the intuitive informational reasons for its existence. Thus, from a quantum informational perspective, the hierarchy of Eq.~(\ref{eq:hierarchy}) is unsurprising. However, exploiting this informational framework and placing the hierarchy in the context of MBL reveals an important consequence: the imbalance - widely used as a quantifier of local memory in MBL - via its informational counterpart $\overline{I_s(t)}$, drastically underestimates memory in MBL systems. In the worst cases (see Fig.~\ref{fig:fig1}) the use of $\overline{I_s(t)}$ erases almost two thirds of the accessible information according to the Holevo quantity.

A final feature of Fig.~\ref{fig:fig1} is that all the four quantities have readily reached a steady-state within very small error bars, even in the midst of the MBLT. This allows us to define time-independent steady-state quantities:
\begin{equation}
    \overline{O}^{(ss)}=\lim_{t \rightarrow \infty} \overline{O(t)}\approx \overline{O(T)}
\end{equation}
where $O$ can be any of our four quantities, namely $\overline{S(t)}$, $\overline{C(t)}$, $\overline{I_c(t)}$ and $\overline{I_s(t)}$, and where we take exponential times of $T = 10^{10}~Jt$ for all results. These steady states can be used for investigating the scaling analysis across the MBL transition point, which is the subject of the next section \ref{sec:scaling} of this paper.

%Thus the only limit relevant to MBL not attained by our analysis is thermodynamic length scales; an intentional choice allowing us to discuss the robustness of these memory quantifiers to the finite-size effect of subsystem dephasing.

%The use of exact diagonalization methods convincingly accesses the dynamical steady-state instead of thermodynamic length scales~\footnote{Our system sizes are not only limited by the conventional `curse of dimensionality', but also by the requirement that we repeat each run $2^l$ times over all possible messages \textit{before} disorder averaging. This isn't an issue here, as we are purposefully investigating how robust these quantities are to finite-size effects, but is a point worth noting.}. This lets us compare the robustness of these quantities to finite-sized effects; informing potential experimental routes to the MBLT through informational quantities, and revealing an interesting result about improving scaling analyses in small systems by artificially dephasing their subsystems - just as a larger systems would do (see our section on scaling analyses and Fig.~\ref{fig:fig3}).

\begin{figure}[!ht]
        \includegraphics[width=\linewidth]{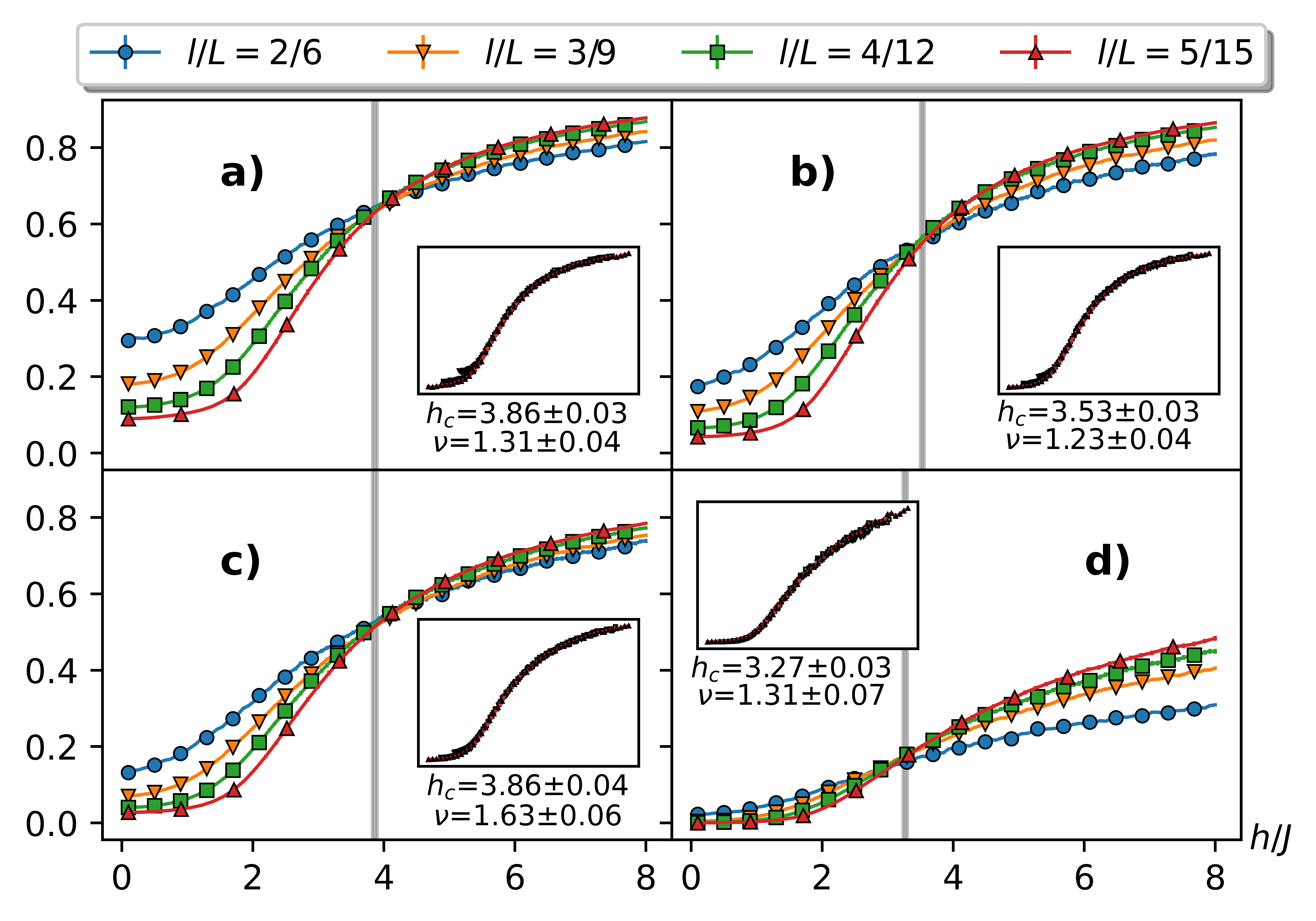}
        \caption{Disorder-averaged steady-state values of (normalized) informational quantities as  a function of disorder strength. The message-to-system size ratio for all panels is $l/L = 1/3$. The panels show: \textbf{(a)} von Neumann entropy \textbf{(b)} Holevo quantity \textbf{(c)} CMI \textbf{(d)} SSMI. Insets show the corresponding data collapse for the three largest systems. Every fifth point is marked for legibility, and error bars shown where visible.}
        \label{fig:fig2}
\end{figure}

\section{Scaling analysis}
\label{sec:scaling}
In our setup, we deal with two explicit lengths, namely message length $l$ and the total system size $L$. Inspired by Ref.~\cite{Chandran2015, Khemani2017}, it is expected that the steady state values follow a finite size anstaz of the form
\begin{equation} \label{eq:ansatz_FSS}
 \overline{O}^{(ss)} = L^{-\zeta/\nu} f(l/L,L^{1/\nu} (h-h_c))
\end{equation}
where $h_c$ is the critical point, $\nu$ and $\zeta$ are critical exponents and   $f(\cdot,\cdot)$ is an unspecified function. 
%Our finite size scaling analysis, discussing below, finds $\zeta = 0$ and thus we will omit it in the next figures. 
By fixing the ratio of $l/L=1/3$, we plot $\overline{O}^{(ss)}$ as a function of $h$ for various system sizes in Figs.~\ref{fig:fig2}(a)-(d) for all the four quantities, respectively. As is clear from the figure, all four quantities show a crossover between the curves of different sizes. The existence of this crossover indicates that $\zeta = 0$ and determines the critical value $h_c$ for each quantity. Interestingly, the critical point $h_c$ determined by the SSMI is far lower than the ones deduced from the other three quantities.

%Interestingly, the critical values estimated from each quantity almost exactly follow the same hierarchy as Eq.~\ref{eq:hierarchy} with the exception of the CMI that yielding a critical value $h_c = 3.86\pm0.04$ far in excess of its coherent counterpart: the Holevo quantity. This counter-inuitive idea: that the Holevo quantity is outperformed by its decohered counterpart, the CMI, in capturing the MBLT is pervasive and motivates the remainder of this letter. %A similar analysis for the ratio $l/L=1/4$ is given in the appendix.   

% do we shorted to SSMI and CMI?}

A more elaborate scaling analysis~\cite{Kjall2014, Sandvik2010}, based on the principle of data collapse, using the ansatz of Eq.~(\ref{eq:ansatz_FSS}) allows us to more precisely determine $h_c$, $\zeta$, and $\nu$. Indeed, our results with variable $\zeta$ (not shown) confirms that $\zeta \approx 0$ for all quantities. Hence, we set $\zeta = 0$ for the rest of the letter. Respective data collapses for each quantity are shown in the insets of Figs.~\ref{fig:fig2}(a)-(d), each yielding slightly different optimal values for $\nu$ and $h_c$. 
The optimal values for $h_c$ and $\nu$ are depicted in Fig.~\ref{fig:fig3}(a). Similar to many previous studies: small system sizes lead to systematic underestimation of the critical value $\nu$, bounded analytically for this system by $\nu \geq 2$ \cite{Harris1974, Chayes1986, Chandran2015}. Additional scaling results are given in Appendix \ref{sec:app-scaling}.

The von Neumann entropy $\nu = 1.31\pm0.04$, the Holevo quantity $\nu = 1.23\pm0.04$, and the SSMI $\nu = 1.31\pm0.07$ all violate the Harris bound to a similar extent. Unexpectedly, all three of these quantities are outperformed in terms of Harris bound violation by the CMI which attains a critical exponent of $\nu = 1.63\pm0.06$, a remarkably high value for such small system sizes. In this sense we can state that the CMI may capture less information than the Holevo quantity (see the hierarchy of Fig.~\ref{fig:fig1}) but it does better capture the MBLT in the sense that it violates the Harris bound to a lesser extent. Finally, the SSMI performs badly both as the worst quantity with which to capture retained information, and as a quantity which violates the Harris bound to a much greater extent than the CMI. It also underestimates the critical $h_c$ with respect to all other quantities. This indicates that the SSMI, and thus the widely used imbalance, are poor scaling quantities which cannot precisely capture properties of the MBLT. As such the SSMI, and by extension the imbalance, are not useful quantifiers of local memory in any arena other than situations where - due to the simple measurements they require - they are the only ones available.

\section{Emulating thermodynamic behavior}
\label{sec:emulate}
The significantly higher values of $\nu$ for the CMI 
may suggest that this quantifier better captures the behaviour of Eq.~\ref{eq:ham} in the thermodynamic limit than the other quantities. Since the CMI is the decohered counterpart of the Holevo quantity, it indicates that throwing away the off-diagonal terms improves scaling results in terms of reducing the violation of the analytic bound $\nu > 2$. For small systems, where these off-diagonal quantum terms persist, we propose that it is beneficial to artificially set them to zero at the end of a simulation, or to only extract diagonal elements in experiment.

\begin{figure}[!ht]
        \includegraphics[width=\linewidth]{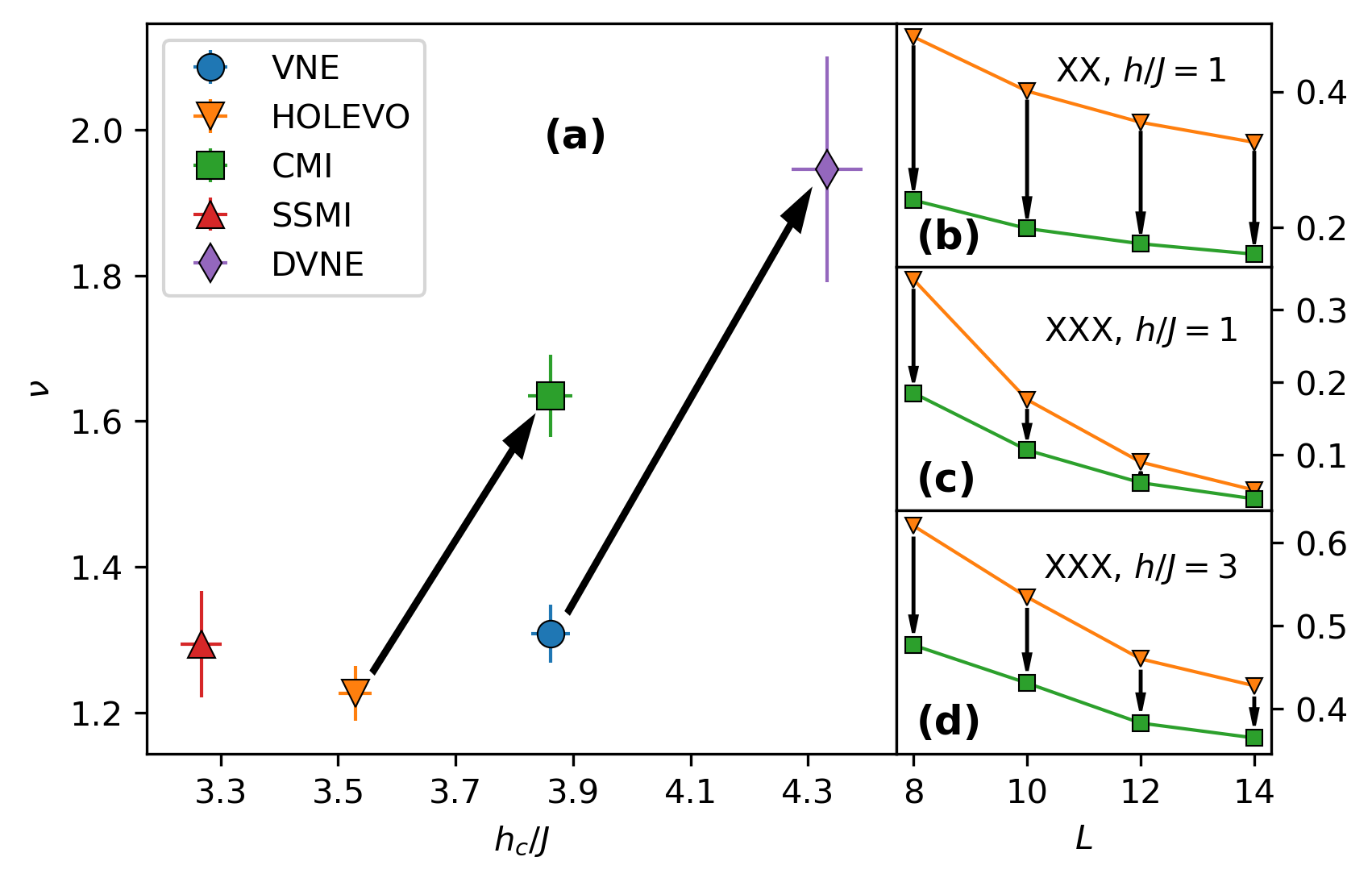}
        \caption{Main panel (a) shows optimal critical values $h_c$ and exponents $\nu$ for all memory quantifiers and the DVNE. Inset panels (b)-(d) compare the Holevo quantity and its decohered counterpart, the CMI, in the XX and Heisenberg (XXX) models. Artificial dephasing is indicated by the arrows.}
        \label{fig:fig3}
\end{figure}

To further support this, we compute the decohered von Neumann entropy (DVNE) by decohering the message register $\rho \to \rho_D$ according to Eq.~(\ref{eq:decohered}) in Eq.~(\ref{eq:vne}). This is essentially the Shannon entropy of diagonal elements in the computational basis, and has been used sparsely in the context of MBL before~\cite{Dalessio2013, Levi2016, Sun2020}. The obtained critical values of $h_c$ and $\nu$ are plotted in Fig.~\ref{fig:fig3}(a). The effect of artificial dephasing according to Eq.~(\ref{eq:decohered}) is shown in Fig.~\ref{fig:fig3} by the black arrows. Remarkably, the von Neumann entropy of artificially dephased quantum states yields a critical value $\nu = 1.95 \pm 0.15$ consistent with the Harris bound~\cite{Harris1974} and closer to large-system analyses of the MBLT. We see similar results in the additional scaling analyses shown in Appendix \ref{sec:app-scaling}.

We conjecture that this may be because artifically decohering the message subsystem imitates the kind of effects that occur in the thermodynamic limit. The decohering of a subsystem in the thermodynamic limit is a direct consequence of interactions, a consequence we investigate by comparing the steady state Holevo quantity and its decohered counterpart, the CMI, directly in interacting and interactionless systems across the MBLT. In Fig.~\ref{fig:fig3}(b) we replace the Heisenberg (XXX) Hamiltonian of Eq.~(\ref{eq:ham}) with the interactionless (XX) Hamiltonian. Even for small disorder strength $h/J=1$, the figure shows slight - if any - convergence between the Holevo quantity and the CMI as $L$ increases. In Fig.~\ref{fig:fig3}(c)-(d) we return to the Heisenberg Hamiltonian. On the small disorder $h/J = 1$ side of the MBLT shown in Fig.~\ref{fig:fig3}(c), the Holevo quantity and the CMI quickly converge as the system size increases. While, close to the transition point $h/J = 3$, we can see from Fig.~\ref{fig:fig3}(c) that the convergence as $L$ increases is still clear, but slowed significantly. From this, we can deduce that the dephasing of the subsystem slows down near the MBL transition point and thus larger system sizes are required to emulate thermodynamic behavior. This explains why wholly quantum quantities may fail to capture the Harris bound $\nu \geq 2$ in small finite systems $L \sim 20$. Essentially, rather than completing this convergence by taking $L$ to a sufficiently large value, our suggestion approximates it by artificially decohering the subsystem instead.

We note here that this procedure enforces a particular understanding of the nature of the MBL state in the thermodynamic limit: as a product of subsystems which eventually dephase. As such we can only conjecture that this procedure emulates the thermodynamic limit, though we support that conjecture with some preliminary evidence (see Fig.~\ref{fig:fig3}(b)-(d)) and argument. What we can state definitively is that this procedure considerably reduces the extent to which the analytic Harris bound is violated in small systems (see Fig.~\ref{fig:fig3}(a)), which in turn is evidence that such an enforcement is plausible.

\section{Conclusions}
\label{sec:conc}
We introduce several quantities for quantification of memory in MBL systems and establish a systematic hierarchical order among them.
Our findings show that the Holevo quantity represents the best quantifier of memory in terms of number of bits of information retained over time, and that the informational version of the widely-used imbalance performs the worst. Furthermore when characterizing the MBL transition using these memory quantifiers we find that, though the Holevo quantity is the best quantifier of memory, the CMI - its decohered counterpart - best captures the critical properties of the MBLT. Motivated by this, we compared the von Neumann entropy of a small subsystem to its decohered variant and discovered that it too outperforms its quantum, coherent, counterpart; yielding a critical exponent consistent with the Harris bound. Our conclusion is that finite-size effects can be mitigated just by deliberately decohering the final state of the message register. This has huge theoretical and experimental advantages: in theory one can drastically improve small-system scaling results and may even be able to emulate the thermodynamic limit by deliberately throwing away information, and in experiment one only needs to measure the diagonal elements of the reduced density matrix instead of extremely demanding state tomography.

\section{Acknowledgements}
\label{sec:ack}
A. B. acknowledges support from the National Key R\&D Program of China (Grant No.2018YFA0306703), the National Science Foundation of China (Grants No. 12050410253 and No. 92065115), and the Ministry of Science and Technology of China (Grant No. QNJ2021167001L). S. B. and A. N. K. acknowledge the
EPSRC (Grant No. EP/R029075/1 for Nonergodic Quantum Manipulation).

\bibliography{refs}

\appendix

\section{The Memory Hierarchy}
\label{sec:app-hierarchy}
Here we derive the first two inequalities in the memory hierarchy $1-\overline{S(t)} \geq \overline{C(t)} \geq \overline{I_c(t)} \geq \overline{I_{s}(t)}$ discussed in the main text. We also discuss a few technical nuances and justifications for our construction of $\overline{I_s(t)}$. We derive the inequalities for a single ensemble of initial probabilities $\{p_k\}$ and final quantum states $\{\rho^{(k)}(t)\}$. Clearly, if it holds for each individual ensemble, it holds for the disorder and environment (channel) average over those ensembles as well. We denote each quantity for a single channel realization using the index $r$, e.g. $\overline{S(t)} = \frac{1}{N_r} \sum_r S^{r}(t)$ where $N_r$ is the number of samples. We suppress this index on the density operators $\rho^{(k)}(t)$ themselves for readability. Thus, the hierarchy we derive here for the largest three quantities is \textit{stricter} than that stated in the main text: $1-S^r(t) \geq C^r(t) \geq I_c^r(t)$.

The first inequality $1-S^r(t) \geq C^r(t)$ trivially follows from our definition of the normalized von Neumann entropy over the ensemble:
\begin{equation}
 S^r(t) = \frac{1}{l}\sum_k p_k S(\rho^{(k)}(t))
\end{equation}
where $S(\rho) = -\text{Tr}\rho\log_2\rho$; and our definition of the Holevo quantity:
\begin{equation}\label{eq:app-holevo}
    C^r(t) = \frac{1}{l}S\left(\sum_k p_k \rho^{(k)}(t)\right)-\frac{1}{l}\sum_k p_k S(\rho^{(k)}(t)).
\end{equation}
By taking the difference of the two quantities $\Delta_{SC}^r(t)$ and noting that the second term in the Holevo quantity is just $S^r(t)$, it follows:
\begin{equation}
    \Delta_{SC}^r(t) = 1-S^r(t)-C^r(t) = 1 - \frac{1}{l}S\left(\sum_k p_k \rho^{(k)}(t)\right)
\end{equation}
where the second term clearly saturates to its maximum value of unity as $\sum_k p_k \rho^{(k)}(t)$ approaches the identity. As such $\Delta_{SC}^r(t) \geq 0$ and the inequality $1-S^r(t) \geq C^r(t)$ follows. This gives us an intuitive understanding of the Holevo quantity as a modification to $1-S^r(t)$; accounting for how much information is \textit{accessible} at time $t$ rather than how useful the $\rho^{(k)}(t)$ are as an instantaneous alphabet.

\begin{figure}[!ht]
        \includegraphics[width=\linewidth]{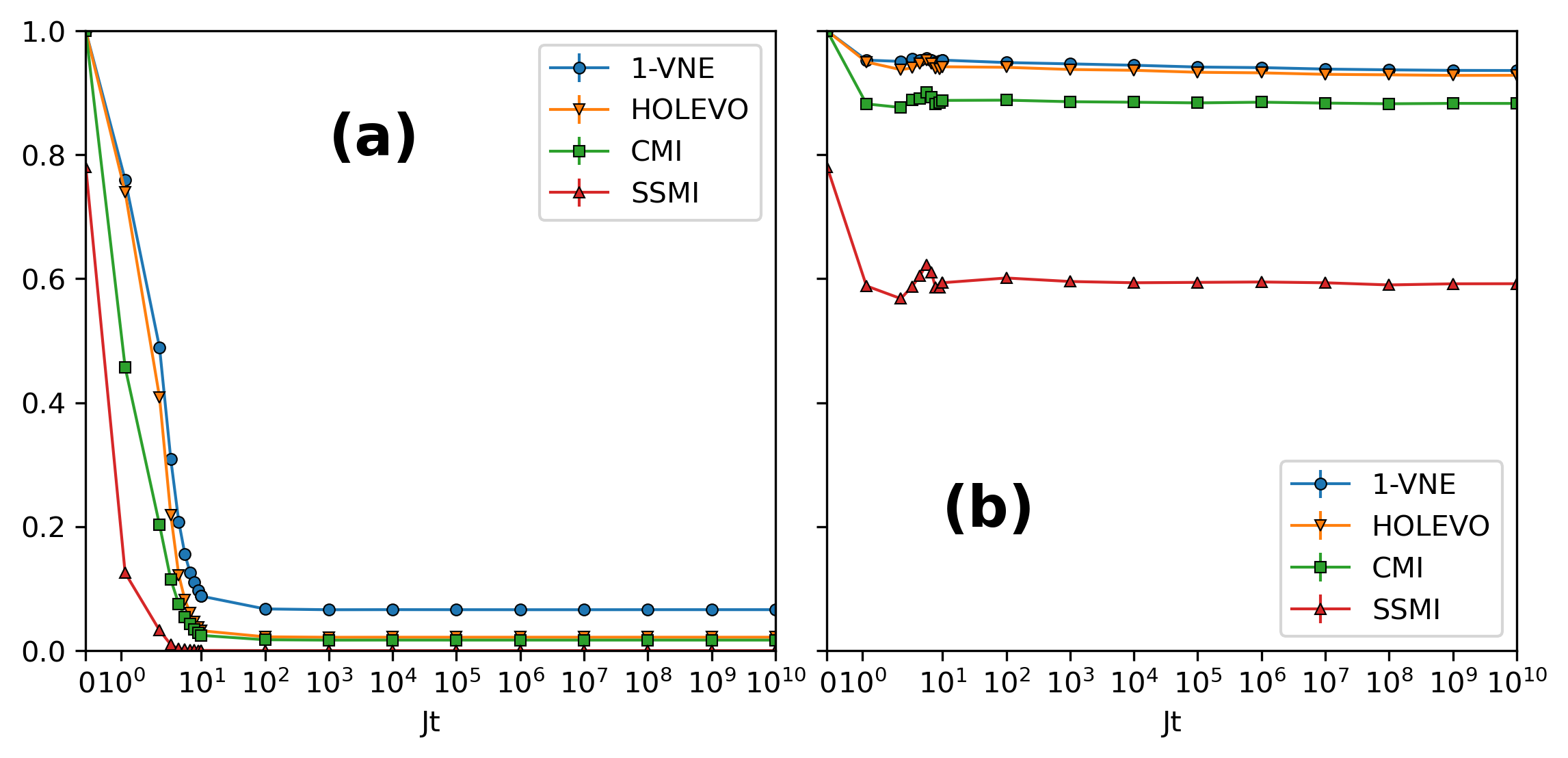}
        \caption{Dynamics of the four informational quantities for $L=16,~l=4$ on extreme sides of the MBLT \textbf{(a)} $h=0.1$, \textbf{(b)} $h=16$, results for the middle of the MBLT ($h=4$) are given in main text. In all cases, a strict hierarchy is seen, and the system convincingly relaxes to a steady-state at exponential times.}
        \label{fig:app-steadystates}
\end{figure}

The derivation of the second inequality is slightly more involved, first requiring us to show that the CMI $I_c^r(t)$ is just a fully decohered version of the Holevo quantity $C^r(t)$. We will carry out the derivation in the computational basis, but in principle any suitable basis can be used instead. In the computational basis, and given the $\{p_k\}$ \textit{a priori}, the joint probability $p(k, k^\prime, t)$ of sending the state $k$ and measuring the state $k^\prime$ at time $t$ is determined entirely by the diagonal elements of $\rho^{(k)}(t)$:
\begin{equation}\label{eq:app-icprobs}
    p(k, k^\prime, t) = p_k \rho_{k^\prime k^\prime}^{(k)}(t)
\end{equation}
we then compute the classical mutual information according to the well-known formula
\begin{equation}\label{eq:app-mutual_c}
    I_c^r(t) = \frac{1}{l}\sum_{kk^\prime} p(k, k^\prime, t)\log_2 \frac{p(k, k^\prime, t)}{ \sum_s p(k, s, t) \sum_{s^\prime} p(s^\prime, k^\prime, t)}
\end{equation}
where the first marginal distribution $\sum_s p(k, s, t)$ resolves trivially to $p_k$ by the unit-trace condition (see Eq.~(\ref{eq:app-icprobs})). The second marginal distribution $\sum_{s^\prime} p(s^\prime, k^\prime, t)$ involves all message states and is generally non-trivial. Evaluating the first marginal distribution, splitting up the logarithm, and cancelling relevant terms yields:
\begin{align*}
    I_c^r(t) = &\frac{1}{l}\sum_{kk^\prime} p_k \rho^{(k)}_{k^\prime k^\prime}(t) \log_2 \rho^{(k)}_{k^\prime k^\prime}(t) \\ &- \frac{1}{l}\sum_{kk^\prime} p_k \rho^{(k)}_{k^\prime k^\prime}(t) \log_2 \sum_{s^\prime} p_{s^\prime} \rho^{(s^\prime)}_{k^\prime k^\prime}(t) 
\end{align*}
we note that, by constructing an operator with the $\rho^{(k)}_{k^\prime k^\prime}(t)$ as eigenvalues, the sums over $k^\prime$ can be replaced with appropriate traces and some terms will take the form of von Neumann entropies. A naturally suitable operator is the the fully decohered (in the computational basis) operator $\rho^{(k)}_D(t)$, defined as the leading diagonal of $\rho^{(k)}(t)$:
\begin{equation} \label{eq:app-decohered}
    \rho^{(k)}(t)~\to~\rho^{(k)}_D(t) = \text{diag}\left(\rho_{11}^{(k)}(t), \rho_{22}^{(k)}(t), \cdots, \rho_{2^l 2^l}^{(k)}(t)\right)
\end{equation}
which trivially has the $\rho^{(k)}_{k^\prime k^\prime}(t)$ as eigenvalues. We can then recast $I_c^r(t)$ in terms of traces as follows:
\begin{align*}
    I_c^r(t) = &\frac{1}{l}\sum_{k} p_k \text{Tr}\left[ \rho_D^{(k)}(t) \log_2 \rho_D^{(k)}(t)\right] \\
    &- \frac{1}{l}\text{Tr}\left[\left(\sum_k p_k \rho_D^{(k)}(t) \right) \log_2 \left(\sum_s p_s \rho_D^{(s)}(t) \right)\right]
\end{align*}
which, when rewritten in terms of the von Neumann entropy $S(\rho) = -\text{Tr}\rho\log_2\rho$, yields the Holevo quantity over decohered states $\rho^{(k)}_D(t)$:
\begin{equation}\label{eq:app-holevo-dephased}
    I_c^r(t) = \frac{1}{l}S\left(\sum_k p_k \rho^{(k)}_D(t)\right)-\frac{1}{l}\sum_k p_k S(\rho^{(k)}_D(t)).
\end{equation}
demonstrating, by analogy to Eq.~(\ref{eq:app-holevo}), that the configurational mutual information as we have constructed it is just the Holevo quantity in the full decoherence limit. Though we have chosen the computational basis for our derivation, any suitable (orthonormal) basis can be used. We can intuit that different choices of basis yield different diagonal elements in the $\rho_D^{(k)}(t)$, and act as effective modifications to the eigenvalues of all the $\rho^{(k)}(t)$ at once. The Holevo quantity is `optimal' in the sense that it uses the true eigenvalues of the $\rho^{(k)}(t)$, equivalent to performing measurements in the eigenbasis of each individual $\rho^{(k)}(t)$.

\begin{figure}[!ht]
        \includegraphics[width=\linewidth]{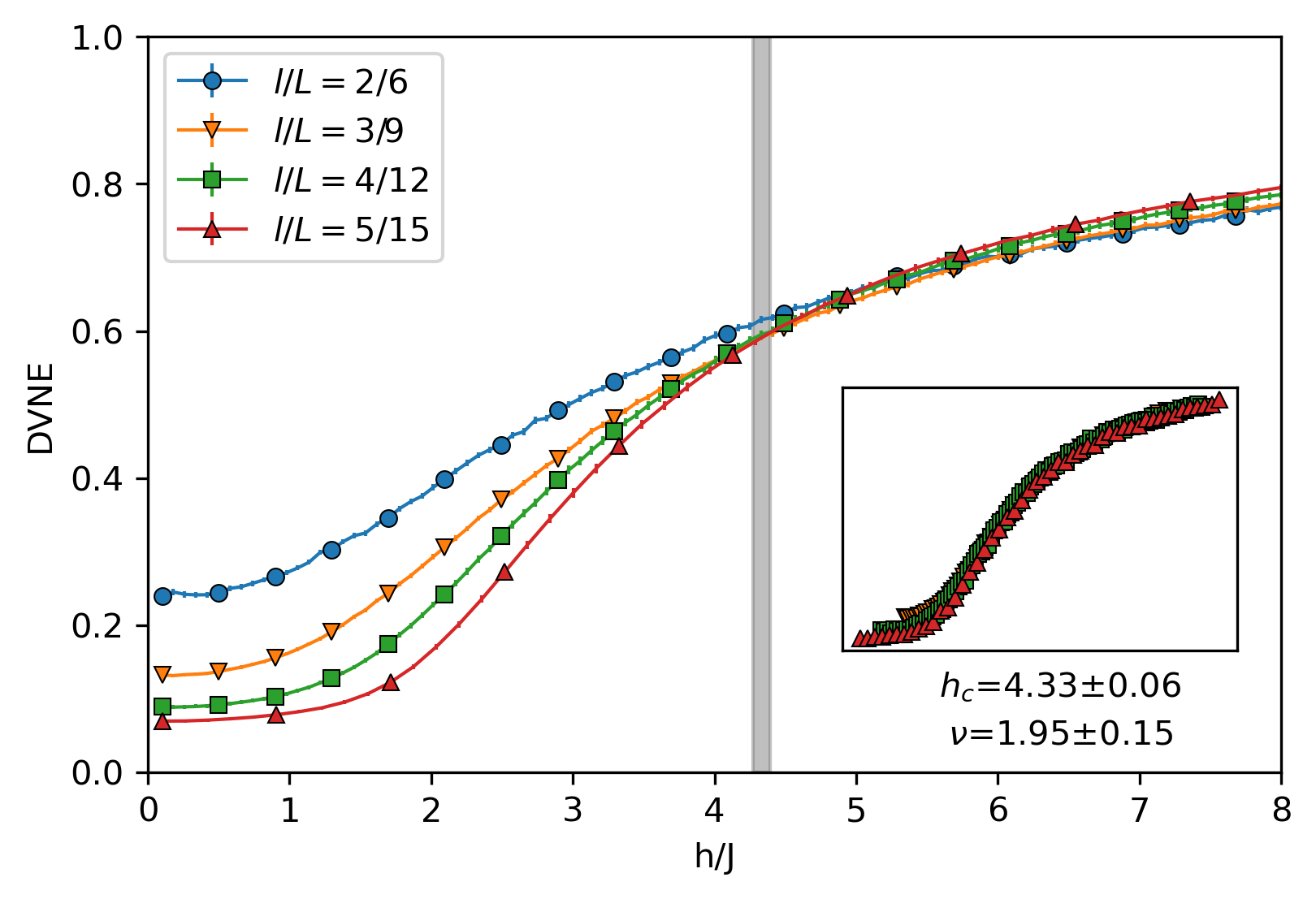}
        \caption{Scaling of the DVNE discussed in the main text (message to system size ratio $l/L = 1/3$). Every fifth point is marked for readability, and error bars shown where visible. The inset shows the corresponding data collapse of the largest three systems. }
        \label{fig:app-dve}
\end{figure}

We now derive the second strict inequality $C^r(t) \geq I_c^r(t)$ by defining their difference:
\begin{equation}
    l\Delta_{CI}^r(t) = lC^r(t) - lI^r_c(t)
\end{equation}
where we have multiplied by $l$ to suppress the normalization constants and avoid notational clutter in the derivation to come. We then insert the Holevo quantity of Eq.~(\ref{eq:app-holevo}) and the dephased Holevo form of the CMI of Eq.~(\ref{eq:app-holevo-dephased}), and fully expand the von Neumman entropies:
\begin{align*}
    l\Delta^r_{CI}(t) = &-\text{Tr}\left[\left(\sum_k p_k \rho^{(k)}(t)\right) \log_2 \left(\sum_k p_k \rho^{(k)}(t)\right) \right]\\
    &+\sum_k p_k \text{Tr}\left[\rho^{(k)}(t) \log_2 \rho^{(k)}(t) \right] \\
    &+\text{Tr}\left[\left(\sum_k p_k \rho^{(k)}_D(t)\right) \log_2 \left(\sum_k p_k \rho^{(k)}_D(t)\right) \right] \\
    &-\sum_k p_k \text{Tr}\left[\rho^{(k)}_D(t) \log_2 \rho^{(k)}_D(t) \right]. 
\end{align*}
We now remark that the operators $\log_2 \rho^{(k)}_D(t)$ and $\log_2 \sum_k p_k \rho^{(k)}_D(t)$ are diagonal by definition, so any matrix multiplication between them and another operator will only involve the diagonal elements of both. Thus, all the $\rho^{(k)}_D(t)$ \textit{outside} those logarithms can be replaced by the corresponding original operators $\rho^{(k)}(t)$. We perform this replacement, collect like terms, and manipulate $l\Delta^r_{CI}(t)$ into a revealing form:
\begin{widetext}
    \begin{align*}
        l\Delta^r_{CI}(t) = &-\text{Tr}\left[\left(\sum_k p_k \rho^{(k)}(t)\right) \left( \log_2 \left(\sum_k p_k \rho^{(k)}(t)\right) - \log_2\left(\sum_k p_k \rho^{(k)}_D(t)\right) \right) \right] \\
        &+\sum_k p_k \text{Tr}\left[\rho^{(k)}(t)\left(\log\rho^{(k)}(t) - \log \rho^{(k)}_D(t) \right)\right]
    \end{align*}
\end{widetext}
in which both terms clearly take the form of quantum relative entropies $S(\rho||\sigma) = \text{Tr}[\rho(\log_2\rho - \log_2\sigma)]$. Explicitly, we can rewrite $\Delta_{CI}^r(t)$ as follows (re-introducing the normalization factor):
\begin{align*}
    \Delta_{CI}^r(t) = &-\frac{1}{l} S\left(\sum_k p_k \rho^{(k)}(t) \right|\left|\sum_k p_k \rho^{(k)}_D(t)\right)\\  &+ \frac{1}{l} \sum_k p_k S\left(\rho^{(k)}(t) \right|\left|\rho^{(k)}_D(t)\right)
\end{align*}
in which, by the joint convexity of the quantum relative entropy, the second term is greater than or equal to the first. As such, $\Delta_{CI}^r(t) \geq 0$ and the Holevo quantity upper bounds the CMI $C^r(t) \geq I_c^r(t)$. We can understand this in two highly intuitive ways: (i) the CMI is `suboptimal' with respect to the Holevo quantity in the sense that it uses a single, fixed, measurement basis (ii) decoherence and measurements cannot increase the information accessible in a system. Numerical results which clearly show this hierarchy in the middle of the many-body localization transition (MBLT) for $h=4$ are given in the main text, whilst results on the extreme sides of the transition ($h=.1$ and $h=16$) are shown in Fig.~\ref{fig:app-steadystates}.

We conclude this section with a brief discussion of the single-site mutual information (SSMI) $\overline{I_s(t)}$ as defined in the main text. We intuit the final inequality $\overline{I_c(t)} \geq \overline{I_s(t)}$ in a similar fashion: the SSMI is constructed using repeated partial traces and projections on the systems used in the construction of the CMI. Such operations cannot increase the information in a system, and so we expect the CMI to systematically upper-bound the SSMI. This has been convincingly verified numerically (see main text). Finally, in our definition of the SSMI in the main text, we average over individual sites and messages in an order that may appear arbitrary: taking the site average at the level of individual probabilities, and the message average at the level of disorder/environment (channel) averaging. Our justifications follow four ideas: (i) an average must be taken, without averaging the protocol sends a single message with probability unity, and as such cannot bear information. This can be verified by inspection of the definition of $\overline{I_s(t)}$ in the main text (ii) For a given site, the rest of the message forms part of the environment (iii) We define the SSMI analogously to the conventional imbalance which averages over each site per realization (iv) our results (not shown) for taking the combined order - with both site and message averaging at the level of individual probabilities - yield no scaling results, and show an overall decrease in informational content with increasing system size; indicating that this decoding protocol is essentially useless.

\begin{figure}[!ht]
        \includegraphics[width=\linewidth]{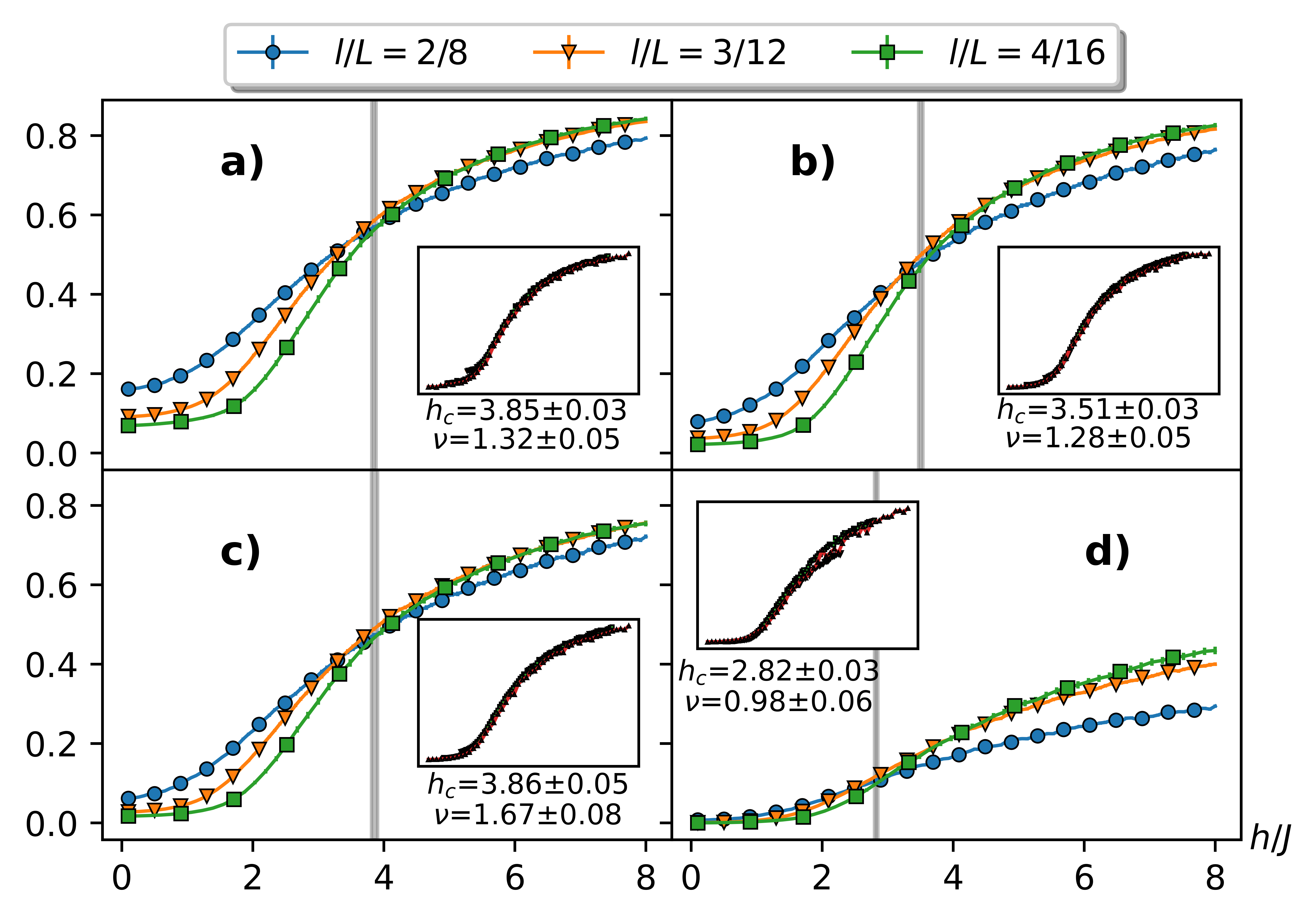}
        \caption{Steady-state values of the four informational quantities \textbf{(a)} von Neumann \textbf{(b)} Holevo quantity \textbf{(c)} CMI \textbf{(d)} SSMI, against disorder for the message to system ratio $l/L = 1/4$. Inserts show corresponding data collapses. Gray regions depict the extracted critical values and their errors.}
        \label{fig:app-l4}
\end{figure}

\section{Additional Scaling Results}
\label{sec:app-scaling}
In this section we provide, for completeness, additional results regarding the characterization of the MBLT. In the main text we discuss the decohered von Neumann entropy (DVNE), which is essentially a Shannon entropy of the diagonal elements of the density matrix in some given basis. The steady-states and corresponding data collapse of the DVNE used in the main text (for $l/L = 1/3$) are given in Fig.~\ref{fig:app-dve}. We can see that - with the exception of the two-site $l=2, L=6$ case - the curves intersect. Critical value $h_c = 4.33\pm0.06$ and exponent $\nu = 1.95 \pm 0.15$ were retrieved from the corresponding data collapse (inset of Fig.~\ref{fig:app-dve}) of the largest three system sizes ($L=9, 12, 15$).

\begin{figure}[!ht]
        \includegraphics[width=\linewidth]{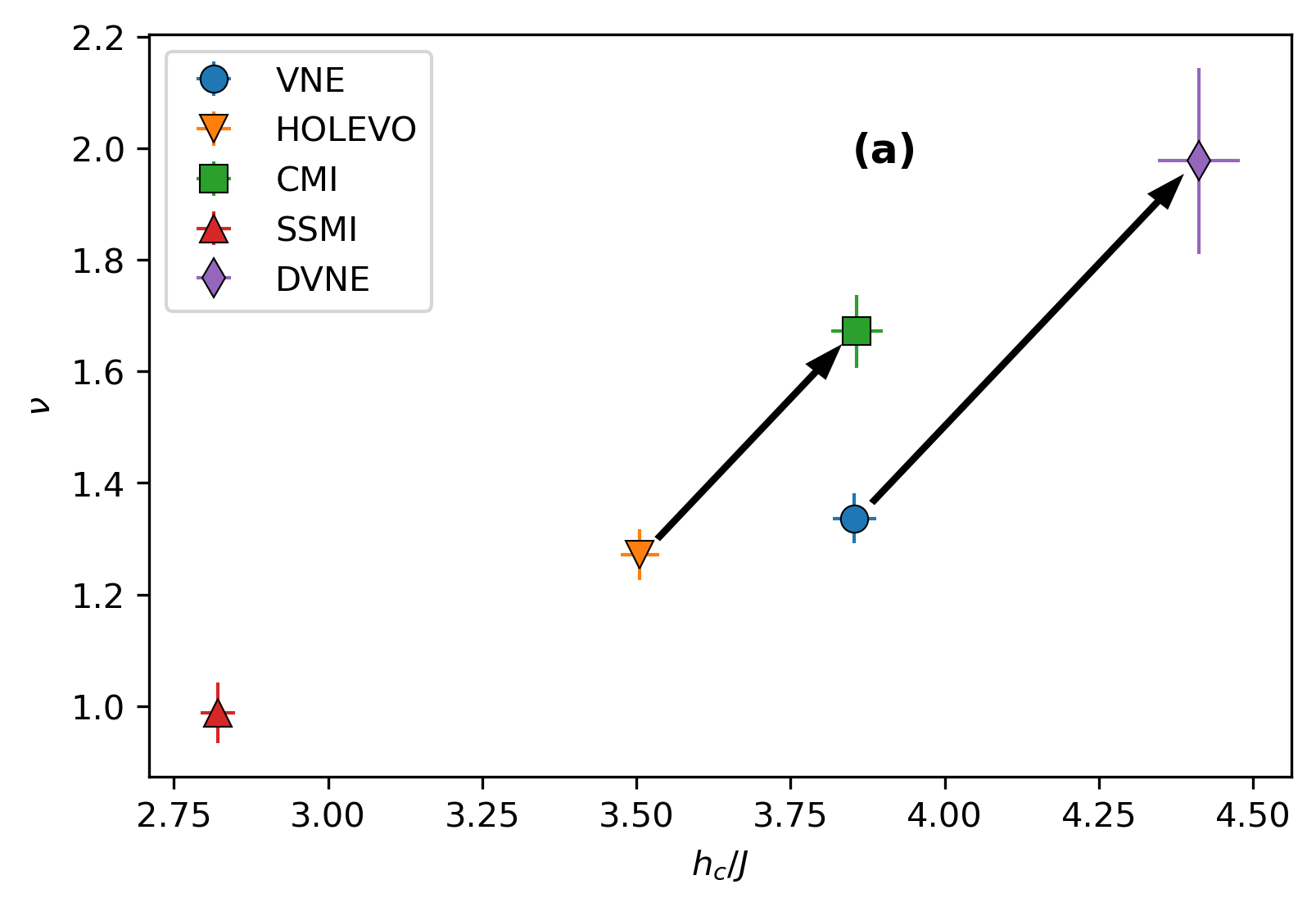}
        \caption{Extracted critical values and exponents for the memory quantifiers for the message to system size ratio $l/L = 1/4$. Black arrows show artificial decohering (see Eq~(\ref{eq:app-decohered}) and main text) which systematically increases the extracted critical values to be closer to large-scale phenomenological studies and analytic bounds. The DVNE is even consistent with the Harris bound $\nu = 1.98 \pm 0.17 \geq 2$.}
        \label{fig:app-l4-crits}
\end{figure}

We also provide the identical analysis as given in the main text, but for the length ratio $l/L = 1/4$ instead. Due mostly to the fact that the numerical procedure by which the Holevo quantity is calculated scales with an additional factor of order $\mathcal{O}(2^l)$, on top of the already $\mathcal{O}(2^L)$ scaling of the underlying Hilbert space, this is limited by computational power to three curves: $L=8, 12, 16$. As such the scalings may suffer from finite-size effects different to, and stronger than, those discussed in the main text and above. The steady-states are shown in Fig.~\ref{fig:app-l4}, and the corresponding critical values and exponents are shown in Fig.~\ref{fig:app-l4-crits}. Despite this, we can see that the decoherence procedure (black arrows) still systematically increases $\nu$ towards the results of analytics and large phenomenological analyses of the MBLT.

\end{document}